\documentclass[conference]{IEEEtran}
\IEEEoverridecommandlockouts
\usepackage{cite}
\usepackage{amsmath,amssymb,amsfonts}
\usepackage{algorithmic}
\usepackage{graphicx}
\usepackage{textcomp}
\usepackage{xcolor}
\def\BibTeX{{\rm B\kern-.05em{\sc i\kern-.025em b}\kern-.08em
    T\kern-.1667em\lower.7ex\hbox{E}\kern-.125emX}}
\begin{document}

\title{An SCMA Receiver for 6G NTN based on Multi-Task Learning\\
}

\author{\IEEEauthorblockN{Bruno De Filippo\IEEEauthorrefmark{1}, Carla Amatetti\IEEEauthorrefmark{1}, Riccardo Campana\IEEEauthorrefmark{1}, Alessandro Guidotti\IEEEauthorrefmark{2}, Alessandro Vanelli-Coralli\IEEEauthorrefmark{1}}
\IEEEauthorblockA{\IEEEauthorrefmark{1}Department of Electrical, Electronic, and Information Engineering (DEI), Univ. of Bologna, Bologna, Italy}
\IEEEauthorblockA{\IEEEauthorrefmark{2}National Inter-University Consortium for Telecommunications (CNIT), Parma, Italy}
\{bruno.defilippo, carla.amatetti2, riccardo.campana7, a.guidotti, alessandro.vanelli\}@unibo.it}

\maketitle

\begin{abstract}
Future 6G networks are envisioned to enhance the user experience in a multitude of different ways. The unification of existing terrestrial networks with non-terrestrial network (NTN) components will provide users with ubiquitous connectivity. Multi-access edge computing (MEC) will enable low-latency services, with computations performed closer to the end users, and distributed learning paradigms. Advanced multiple access schemes, such as sparse code multiple access (SCMA), can be employed to efficiently move data from edge nodes to spaceborne MEC servers. However, the non-orthogonal nature of SCMA results in interference, limiting the effectiveness of traditional SCMA receivers. Hence, NTN links should be protected with robust channel codes, significantly reducing the uplink throughput. Thus, we investigate the application of artificial intelligence (AI) to SCMA receivers for 6G NTNs. We train an AI model with multi-task learning to optimally separate and receive superimposed SCMA signals. Through link level simulations, we evaluate the block error rate (BLER) and the aggregated theoretical throughput achieved by the AI model as a function of the received energy per bit over noise power spectral density ratio (Eb/N0). We show that the proposed receiver achieves a target 10\% BLER with 3.5dB lower Eb/N0 with respect to the benchmark algorithm. We conclude the assessment discussing the complexity-related challenges to the implementation of the AI model on board of a low earth orbit satellite.
\end{abstract}

\begin{IEEEkeywords}
Non-Orthogonal Multiple Access, Deep Learning, Multi-Task Learning, Non-Terrestrial Networks
\end{IEEEkeywords}

\section{Introduction}
\noindent Non-terrestrial networks (NTNs) represent a crucial component of modern communication infrastructures, becoming a key technology of the future 6th generation (6G) of cellular communication standards \cite{bib:ntn3GPP}. Interconnected satellites at different altitudes provide extensive coverage, especially in remote and underserved regions, integrating with terrestrial networks (TNs) to reach ubiquitous connectivity. In parallel, edge computing frameworks, such as multi-access edge computing (MEC), have emerged as enablers for low latency services and distributed learning paradigms. As an example, nodes distributed on ground can collect local data and train a local version of an artificial intelligence (AI) model, sharing only the model updates with the MEC server in a federated learning fashion \cite{bib:MEC}. However, the communication link can be a bottleneck to such use cases, as too many users may request access to the network simultaneously. Non-orthogonal multiple access (NOMA) techniques have been greatly investigated as a mean to enhance the uplink spectral efficiency and accommodate massive connectivity demands. Among such techniques, the codebook-based sparse code multiple access (SCMA) provides a good balance between performance and complexity. SCMA enables multiple users to sparsely share the same frequency resources, significantly boosting the spectrum efficiency while managing the generated interference. On the opposite, as the transmissions originating from the same beam exhibit similar received power levels, power-domain NOMA approaches appear as less promising for NTNs \cite{bib:confCoherent}. In this framework, the interplay between NTNs, SCMA, and MEC generates a strong synergy. The use of SCMA in NTNs ensures that a vast number of transmissions, \textit{e.g.}, collected data or AI model updates, from ground devices (GDs) are supported. LEO satellites, functioning as a MEC servers, processes this data at the edge, thereby enabling rapid responses and decision-making capabilities \cite{bib:MEC_NTN}. However, the NTN propagation channel poses significant challenges to the communication links, which are typically operating at extremely low received power levels. In order to improve the link budget, more complex ground terminals should be implemented, \textit{e.g.}, employing highly directive and steerable antennas. To keep the terminal complexity at a minimum, the link can be enhanced at the receiver side, \textit{e.g.}, using AI techniques such as neural networks (NNs).
\noindent SCMA has not been widely investigated in the context of NTN. In \cite{bib:confCoherent}, a cross-correlation-based Doppler estimation scheme was proposed for low earth orbit (LEO)-based NTNs. The authors showed that a 10\% pilot overhead results in a performance gap with respect to ideal channel estimation of just 1 dB. A contention-resolution procedure was developed in \cite{bib:satColliding} to mitigate the effects of codebook collisions in NTN SCMA grant-free systems, resulting in a bit error rate (BER) loss of just 1dB of energy per bit over noise power spectral density ratio (Eb/N0) with respect to non-colliding SCMA transmissions. With the aid of AI, an autoencoder-based SCMA NTN system was presented in \cite{bib:dlNTN}, implementing the SCMA encoder and decoder with convolutional layers and the propagation channel with a wasserstein generative adversarial network (WGAN). The proposed algorithm achieved a BER gain of 6dB of Eb/N0 in an additive white gaussian noise (AWGN) channel with respect to the traditional SCMA scheme. AI-based SCMA receivers have also been proposed for TNs, \textit{e.g.}, in \cite{bib:jointDemDec}, where a modular NN for the joint demodulation of SCMA symbols and decoding of non-binary polar-coded bits was presented. The model achieved a 2.5 dB Eb/N0 gain in block error rate (BLER) over the benchmark under AWGN, while also providing complexity reductions. On the other hand, the authors in \cite{bib:SCMAdist} presented a NN to limit the search space of the message passing algorithm (MPA), reducing the overall computational complexity without significant impact on the BER. Generative AI approaches were also proposed in \cite{bib:AEfirst, bib:WGANsecond, bib:residual} for various TN channel conditions.

The literature has shown that NNs are promising algorithms to improve SCMA receivers. The sparse patterns of SCMA signals can be identified by AI models, moving from iterative receivers to highly parallelizable algorithms. However, non-generative approaches to SCMA in NTN have not been evaluated yet. While SCMA can provide the desired spectral efficiency boost to 6G NTNs, it is also imperative to reduce the Eb/N0 levels required to achieve satisfactory link performance. Thus, we here adopt multi-task learning (MTL) to let each section of the NN adapt to a specific SCMA codebook. The main contribution of this work are the following:
\begin{itemize}
    \item We present an MTL-based SCMA receiver optimized for NTN, requiring no additional complexity at the GDs;
    \item We evaluate the proposed receiver in an end-to-end channel-coded NTN link, taking into consideration the challenges related to the implementation of an AI-based receiver on board of a satellite.
\end{itemize}

\section{System model}
\begin{figure}[t]
    \centerline{\includegraphics[width=8cm]{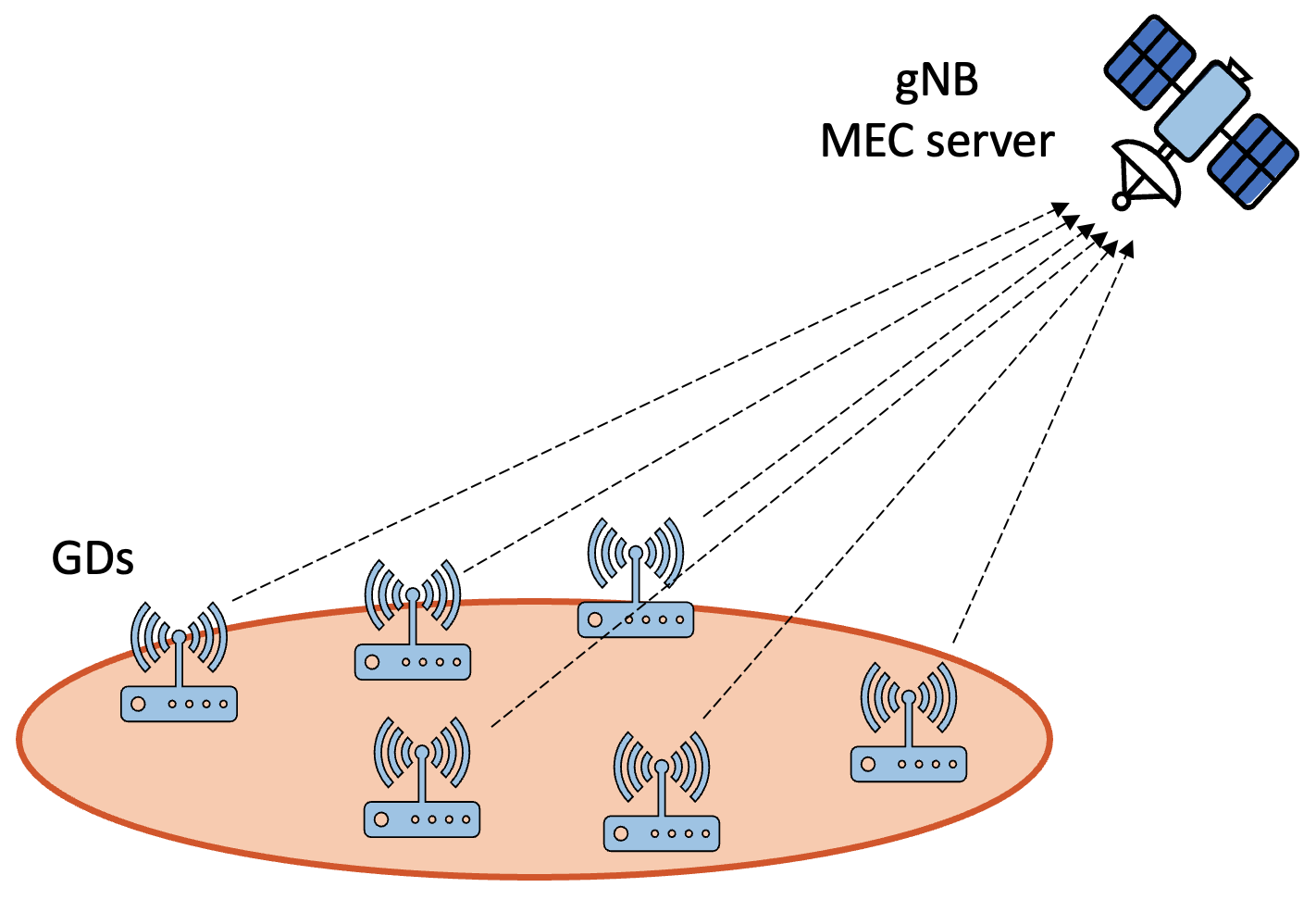}}
    \caption{Considered system model for SCMA-based MEC in NTN.}
    \label{fig:system_model}
\end{figure}
\begin{figure}[t]
    \centerline{\includegraphics[width=8cm]{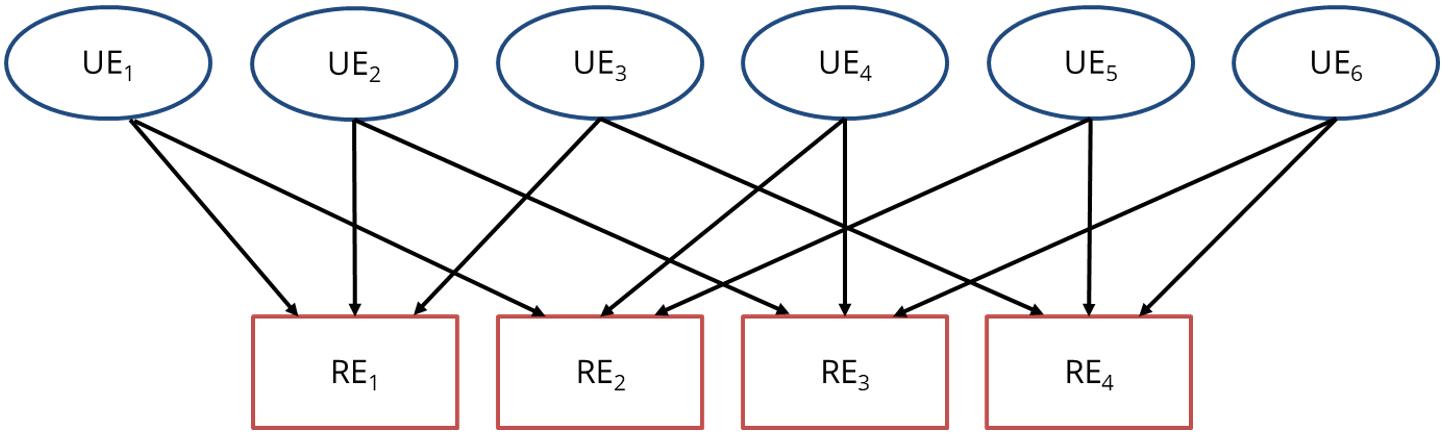}}
    \caption{SCMA factor graph.}
    \label{fig:FG}
\end{figure}

\noindent We consider a LEO satellite providing connectivity to a set of GDs within its coverage area (Figure \ref{fig:system_model}). In our NTN-based MEC framework, the satellite payload carries the MEC server. The GDs act as edge nodes that collect, pre-process, and transmit data to it, \textit{e.g.}, model updates for federated learning, or sensor data for internet of things applications. We assume that the gNode B (gNB), which is co-located with the MEC server on board of the satellite, schedules $J$ GDs in connected mode for uplink SCMA transmission over a set of $K < J$ shared frequency resources. 5G NR physical resource blocks (PRBs), each comprising $N_{SC}^{PRB}=12$ orthogonal frequency division multiple access (OFDMA) subcarriers, are allocated to GDs; thus, we here consider $K$ PRBs as frequency resources. The data payload of the $i$-th GD is mapped to one or more 5G transport blocks (TBs) $\mathbf{b}_i$ of size $N_{TBS}$ bits. Each TB is encoded with a low density parity check (LDPC) channel encoder with code rate $R_c$, resulting in a vector of $N_{bits}$ coded bits $\mathbf{b}_i^{(c)}$. An SCMA encoder and modulator is implemented based on the chosen SCMA codebook $\Gamma_i(\cdot)$, mapping each group of $m$ coded bits in $\mathbf{b}_i^{(c)}$ to one of $M=2^m$ complex codewords of length $K$. Such mapping is repeated for each of the $N_{SC}^{PRB}$ groups of $K$ subcarriers. The SCMA codebooks are designed to minimize the overlap between the codewords of different GDs, allowing an increased amount of GDs to be scheduled over the same set of subcarriers with respect to OFDMA. In particular, each codeword has $d_v$ non-zero elements, and the number of codewords of different GDs overlapping over the same resource is limited to $d_f$. Assuming $K=4$, $J=6$, $d_v=2$ (\textit{i.e.}, two PRBs are allocated to each GD), and $d_f=3$ (\textit{i.e.}, the same PRB is allocated to three GDs), the factor graph (FG) in Figure \ref{fig:FG} can be obtained. The resulting overloading factor with respect to orthogonal transmissions (\textit{i.e.}, one user per PRB) is $\frac{J}{K} = 150\%$.
\noindent At the $i$-th transmitter, the generated SCMA codewords $\mathbf{x}_i = \Gamma_i(\mathbf{b}_i^{(c)})$ are mapped over the $N_{sc} = 12\cdot K$ shared subcarriers, repeating the SCMA FG pattern over the available bandwidth and covering a number of symbol times $N_{sym}$. The resulting grid $\mathbf{X}_{i}$ of size $(12\cdot K, N_{sym})$ is used to generate the cyclic prefix orthogonal frequency division multiplexing (CP-OFDM) waveform, which is then transmitted at carrier frequency $f_c$. Assuming line of sight with the satellite, each signal encounters a different propagation channel, represented by the following matrix $\mathbf{H}$ of single tap channel coefficients:
\begin{equation}\label{eqn:channel_coeff}
    \left[\mathbf{H}\right]_{i,t}=G_{i,t}^{(tx)}G_{i,t}^{(rx)}\sqrt{L_{i,t}}\frac{\lambda}{4\pi d_{i,t}}e^{-j\frac{2\pi}{\lambda}d_{i,t}},
\end{equation}
where $\left[\mathbf{H}\right]_{i,t}$ indicates the element at row $i$ and column $t$ of $\mathbf{H}$ (\textit{i.e.}, the channel coefficient of the $i$-th link at symbol time $t$), $G_{i,t}^{(tx)}$ and $G_{i,t}^{(rx)}$ represent the transmission and reception gain, respectively, $d_{i,t}$ represents the slant range, $\lambda = \frac{c}{f_c}$ is the carrier wavelength (with $c$ being the speed of light), and $L_{i,t}$ contains all of the power losses in addition to the free space path loss, \textit{i.e.}, atmospheric, scintillation, and shadowing losses. At the receiver, the signals are superimposed one with the other and summed with AWGN with power spectral density $N_0$. After OFDMA demapping, the received grid at symbol time $t$ can be represented as follows:
\begin{equation}\label{eqn:rx_signal}
    \mathbf{Y} = \sum_{i=1}^J \left[\mathbf{H}\right]_{i,:} \cdot \mathbf{X}_i + \mathbf{N},
\end{equation}
where $\mathbf{N}$ represents the matrix of AWGN samples that affects the received OFDMA grid. To separate the SCMA symbols of different GDs over the $t$-th symbol time and retrieve the $J$ sequences of coded bits $\{\mathbf{b}_{i}^{(c)}\}_{i=1}^J$, a variant of the MPA is typically implemented at the receiver. These algorithms progressively refine the coded bit log-likelihood ratios (LLRs) corresponding to each SCMA transmission, iteratively computing and exchanging messages between factor nodes (representing the $K$ frequency resources) and variable nodes (representing the $J$ GDs) of the SCMA FG. For a more detailed description of the most popular MPA variants, the reader can refer to \cite{bib:scmaComplexity}. Clearly, the process must be repeated $N_{SC}^{PRB}$ times to cover the $K$ PRBs and for each symbol time encompassing the transmission of the TB. Once the maximum number of iterations is reached and the entire OFDMA grid has been processed, the estimated LLR sequence related to the $i$-th GD $\mathbf{LLR}_i$ is channel decoded, obtaining the corresponding received TB $\hat{\mathbf{b}}_i$. The data is then forwarded to the upper layers, after which the GDs' payloads can finally be processed by the MEC server.
\noindent Given the iterative nature of MPA, the computation of the LLRs corresponding to a received symbol vector at symbol time $t$, $\left[\mathbf{Y}\right]_{:,t}$, requires long sequential operations that cannot be easily parallelized. Furthermore, variants of the MPA with lower computational complexity, such as the Log-MPA or the Max-log-MPA, are typically preferred over the base algorithm, leading to suboptimal performance. For this reason, we opt for a DL-based receiver to perform SCMA decoding and demodulation. Indeed, DL algorithms can provide significant gains by approximating with matricial operations the optimal SCMA receiver.

\section{Proposed AI-based SCMA receiver}\label{sec:model}
\begin{figure}[t]
    \centerline{\includegraphics[width=7cm]{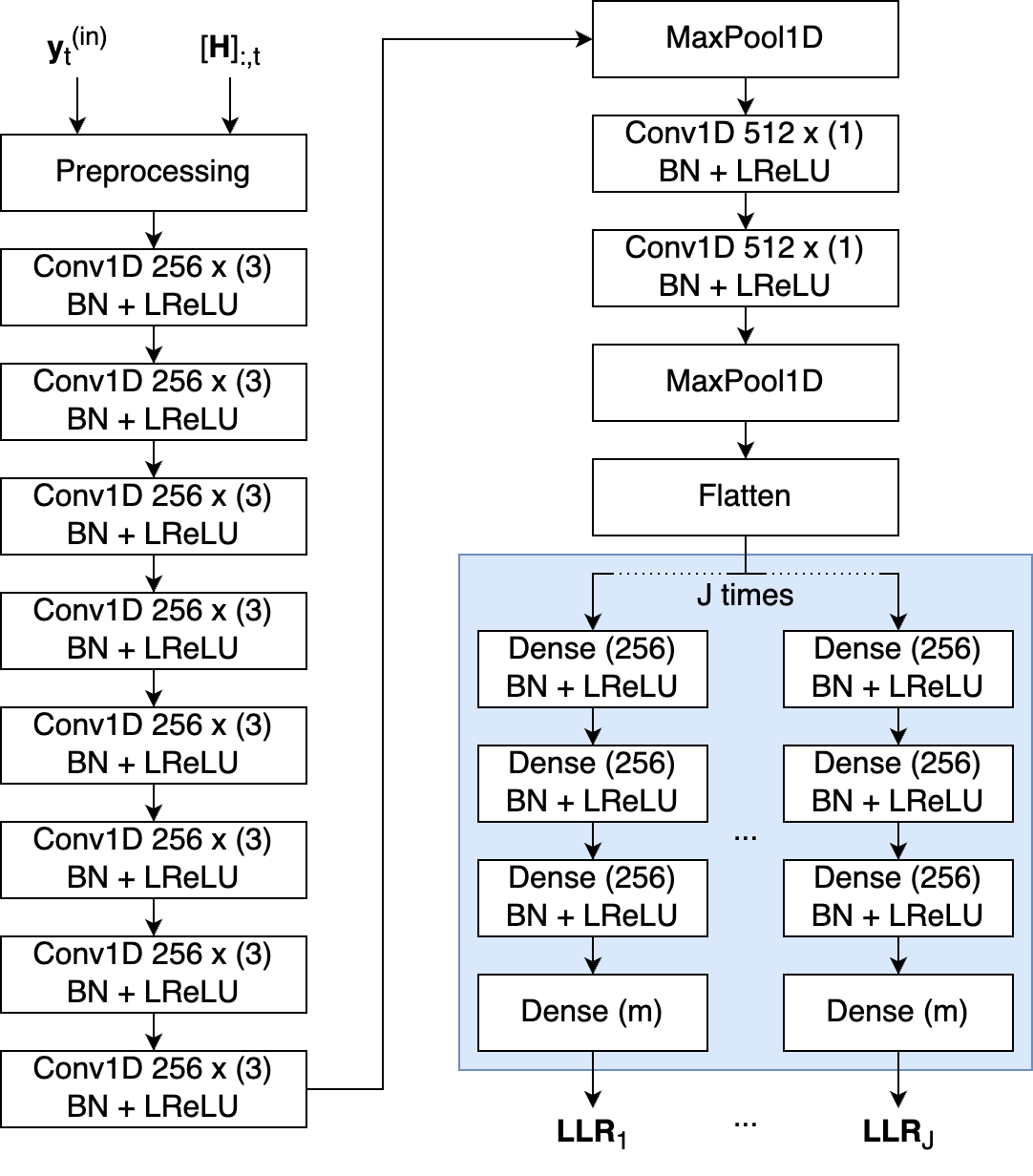}}
    \caption{Diagram of the proposed CNN-based SCMA receiver.}
    \label{fig:NNscheme}
\end{figure}

\begin{figure}[t]
    \centerline{\includegraphics[width=7cm]{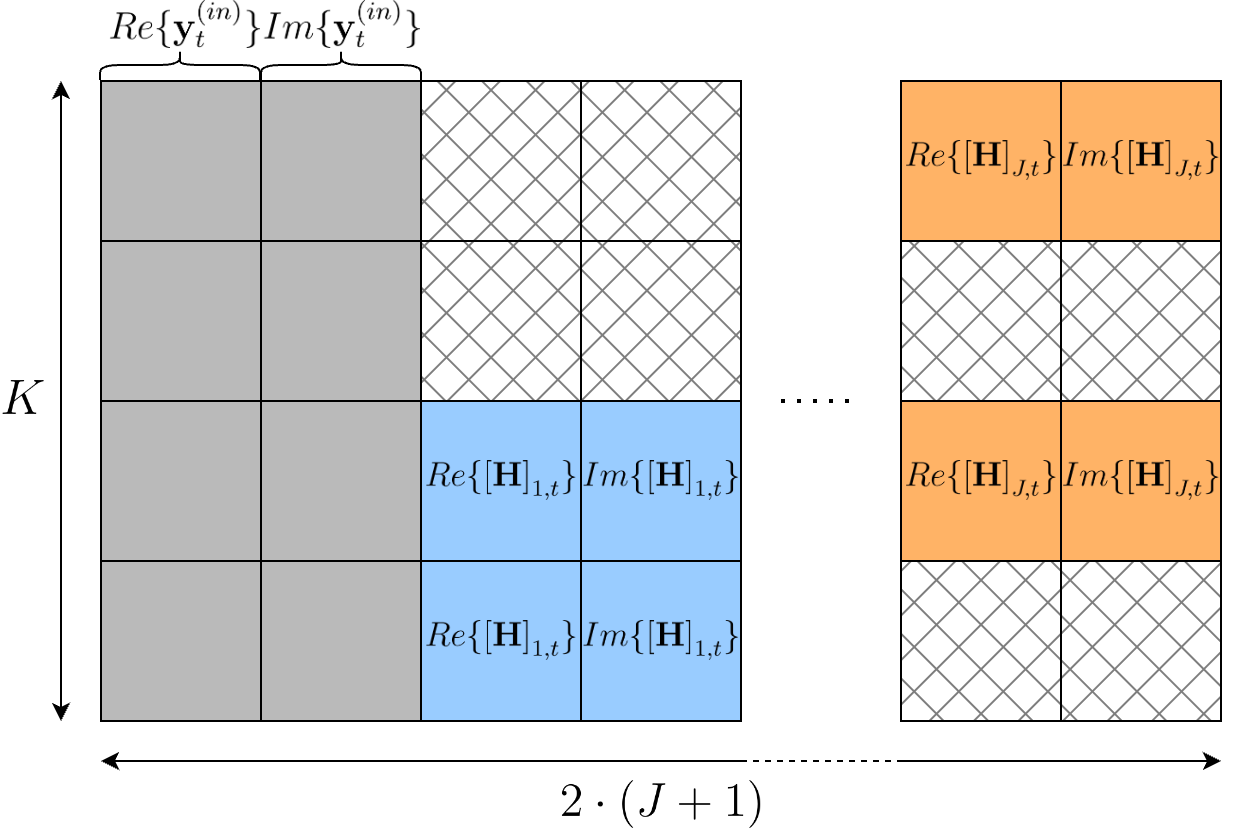}}
    \caption{NN input after the pre-processing reshaping step.}
    \label{fig:preprocess}
\end{figure}

\noindent The task of the proposed receiver consists in estimating $m\cdot J$ coded bit LLRs for each set of $K$ subcarriers at time symbol $t$, given the corresponding complex vectors of $K$ received symbols $\mathbf{y}_t^{(in)}\subset \left[\mathbf{Y}\right]_{:,t}$ and $J$ channel coefficients $\left[\mathbf{H}\right]_{:,t}$. The DL model proposed in this work is a convolutional NN (CNN), aiming at extracting the spatial patterns introduced by the SCMA codebook with convolutional layers. MTL is also employed to train layers that specialize on different tasks, \textit{i.e.}, decoding different SCMA codebooks, based on the same features extracted by common layers. The structure of the CNN is reported in Figure \ref{fig:NNscheme}, where the task-specific layers are highlighted in light blue.
\noindent To obtain an image-like input data shape, we first pre-process the user channel coefficients, which are assumed to be known at the receiver, together with the received grid in a similar fashion to \cite{bib:DeepRx}. In particular, each example within a minibatch consists of a real-valued matrix $\mathbf{A}$ of shape $\left(K,\text{ }2\cdot(J+1)\right)$, initialized with zeros. The first and second columns are filled with the real and imaginary parts of the considered $K$ symbols within the received grid, $\mathbf{y}_t^{(in)}$, respectively. Then, $\left[\mathbf{A}\right]_{k, (2*i+1)}$ and $\left[\mathbf{A}\right]_{k, (2*i+2)}$ are filled with $Re\{\left[\mathbf{H}\right]_{i,t}\}$ and $Im\{\left[\mathbf{H}\right]_{i,t}\}$, respectively, if $\Gamma_i(\cdot)$ foresees a transmission on subcarrier $k$. Such operation embeds the SCMA codebook in the NN input, improving the training process. Figure \ref{fig:preprocess} represents the obtained matrix considering the codebook reported in \cite{bib:codebook}.
After collecting $N_e$ examples, the resulting input minibatch is a tensor of shape $(N_e,\text{ }K,\text{ }2\cdot(J+1))$, where the first dimension refers to the minibatches, the second is a spatial dimension related to the subcarrier, and the third is the channels dimension. Before being fed to the CNN, the data in each channel is separately standardized using first and second order statistics, which have been previously estimated on a generated minibatch. To extract and process spatial features from the frequency domain, the input tensor is fed to a set of eight 1-dimensional convolutional (Conv1D) layers with 256 kernels of length 3, employing "same" padding to maintain the length of the spatial/frequency dimension. 
Considering an $L \times D$ padded input matrix $\mathbf{I}$, an $N^{(f)} \times W^{(f)} \times D$ optimized weight tensor $\underline{\mathbf{W}}$ (with $W^{(f)}$ odd), and an $L \times N^{(f)}$ optimized bias matrix $\mathbf{B}$, a Conv1D layer with "same" padding computes an $L \times N^{(f)}$ output matrix $\mathbf{O}$ as follows:
\begin{equation}\label{eqn:Conv1D}
    \left[ \mathbf{O}\right]_{:,n} = \left[ \mathbf{B}\right]_{:,n} + \sum_{d=1}^D \left[ \mathbf{I}\right]_{:,d} * \left[ \underline{\mathbf{W}}\right]_{n,:,d},
\end{equation}
where the symbol $*$ represents the cross-correlation operator. Each Conv1D layer is followed by a batch normalization (BN) layer and a leaky rectified linear unit (LReLU) activation function:
\begin{equation}\label{eqn:LReLU}
    LReLU(x) = \begin{cases}
        x & \text{if } x > 0 \\
        0.3\cdot x & \text{otherwise}
    \end{cases}
\end{equation}
After the first eight layers, a MaxPooling layer is employed, halving the tensor dimension by choosing the maximum between each group of two elements over the spatial dimension. Two Conv1D layers with 512 kernels of size 1 and "same" padding are then used to refine the extracted features. The common features extraction section of the model terminates with another MaxPooling layer, flattening the tensor to a vector with 512 elements. The extracted features are fed to J task-specific chains with identical structure, where the $i$-th chain specializes on the estimation of the $m$ bit LLRs of GD $i$. Each chain is composed of 3 dense layers with 256 neurons, each activated with LReLU and followed by a BN layer. Each chain is completed by a dense layer with $m$ neurons, leading to a model output of length $m\cdot J$. The receiver should perform a typical classification task, in which each output $\hat{l}_u$, $u \in [1, m\cdot J]$, is activated by the sigmoid function $\sigma(\hat{l}_u)$ to obtain the probability of the $u$-th output bit being 1, $\hat{b}_u = \sigma(\hat{l}_u)=(1+e^{-\hat{l}_u})^{-1}$. Clearly, this representation carries the same information as the $u$-th bit LLR, which actually coincide with the non-activated value $\hat{l}_u$ of the output tensor:
\begin{equation}\label{eqn:LLR}
    LLR_u = ln\left(\frac{Pr\{b_u = 1 | \hat{b}_u\}}{Pr\{b_u = 0 | \hat{b}_u\}}\right) = ln\left(\frac{\hat{b}_u}{1-\hat{b}_u}\right) = \hat{l}_u,
\end{equation}
where $b_u$ represents the true value of the $u$-th output bit. For this reason, we move $\sigma(\cdot)$ to the loss function, resulting in the binary cross-entropy with logits (L-BCE):
\begin{equation}\label{eqn:L-BCE}
    \begin{split}
        L-BCE = & \frac{1}{m\cdot J} \sum_{u = 1}^{m\cdot J} L-BCE_u\text{, with } \\
        L-BCE_u = & - b_u\cdot ln\left(\sigma\left(\hat{l}_u\right)\right) + \\
                  & - (1 - b_u)\cdot ln\left(1 - \sigma\left(\hat{l}_u\right)\right).
    \end{split}
\end{equation}
With little mathematical manipulation, recalling the formulation of $\sigma(\hat{l}_u)$, the $L-BCE_u$ term can be rewritten as:
\begin{equation}
    L-BCE_u =
        \begin{cases}
            ln\left(1 + e^{\hat{l}_u}\right) & \text{ if }b_u = 0 \\
            ln\left(1 + e^{-\hat{l}_u}\right) & \text{ if }b_u = 1
        \end{cases}
\end{equation}
which corresponds to the softplus function $f_{sp}(a) = ln(1 + e^a)$ computed in $a = \pm \hat{l}_u$ based on the value of the true label $b_u$. Thus, training the CNN with the L-BCE loss results in the vector of $m$ LLRs associated to the $i$-th GD, $\mathbf{LLR}_i$, to be estimated at the output of the $i$-th neural chain.

\section{Results}
\setlength{\tabcolsep}{5pt}
\renewcommand{\arraystretch}{1.1}
\begin{table}
    \centering
    \caption{Simulation parameters}
    \label{tab:parameters}
    \begin{tabular}{|c|c|}
        \hline
        \textbf{Parameter} & \textbf{Value} \\
        \hline
        Satellite altitude & 600 km \\
        \hline
        Carrier frequency & $f_c=2$ GHz \\
        \hline
        SCMA codebook & As in \cite{bib:codebook} \\
        \hline
        Number of Log-MPA iterations & $N_{iter}=10$ \\
        \hline
        5G numerology & $\mu = 0$ \\
        \hline
        Code rate & $R_c = \{0.588, 0.188\}$ \\
        \hline
        TBS & $N_{TBS} = \{168, 48\}$ bits \\
        \hline
        Coded bits per TBS & $N_{TBS}^{(c)} = 288$ bits\\
        \hline
        Eb/N0 range & $[-15,10]$ \\
        \hline
        Monte Carlo Iterations & $N_{MC} = 10^4$ \\
        \hline
        Channel model & Based on \cite{bib:3gpp_tr_38.811} \\
        \hline
    \end{tabular}
\end{table}
\noindent In this section, we present a link level performance analysis of the proposed AI algorithm. Table \ref{tab:parameters} provides an overview of the simulation parameters. First, a train and test dataset of channel coefficients was obtained by simulating the orbital movement of a LEO satellite at 600km of altitude on the MATLAB computing environment. Then, the CNN presented in Section \ref{sec:model} was implemented using Python with the TensorFlow library \cite{bib:tensorflow}. A data generator was developed, where user bits are continuously randomly generated, mapped to SCMA codewords, and superimposed one with each other and AWGN considering the training channel coefficients dataset. The model was trained using the popular Adam optimizer, setting the maximum number of epochs to 2000. Each epoch consisted of 256 minibatches of data, each comprising of $N_e=3000$ superimposed SCMA symbols and the corresponding bits as input data and labels, respectively. To improve the learning process, the learning rate, initially set to $10^{-3}$, was reduced by a factor 10 every time a L-BCE improvement was not observed for 50 training epochs. Early stopping was also implemented to interrupt the training process after 200 epochs from the last observed loss decrease. Once trained, the model was imported in MATLAB in a complete end-to-end link-level simulation, including LDPC encoding/decoding and OFDMA waveform generation, based on the 5G Toolbox. We ran a Monte Carlo simulation with $N_{MC}=10^4$ iterations, during which each of the $J=6$ GDs was generating a TB of either $N_{TBS}=168$ bits ($R_c = 0.588$) or $N_{TBS}=48$ bits ($R_c = 0.188$). In both cases, after rate matching the coded TB was composed of $N_{TBS}^{(c)}=288$ bits. We first assessed the BLER as the average number of correctly decoded TBs averaged over the Monte Carlo iterations and the GDs, without hybrid automatic repeat request. Such metric is critical to evaluate the effectiveness of the proposed receiver, reporting the Eb/N0 level required to achieve a target error rate on the transmitted TBs.
\begin{figure}[t]
    \centerline{\includegraphics[width=7.5cm]{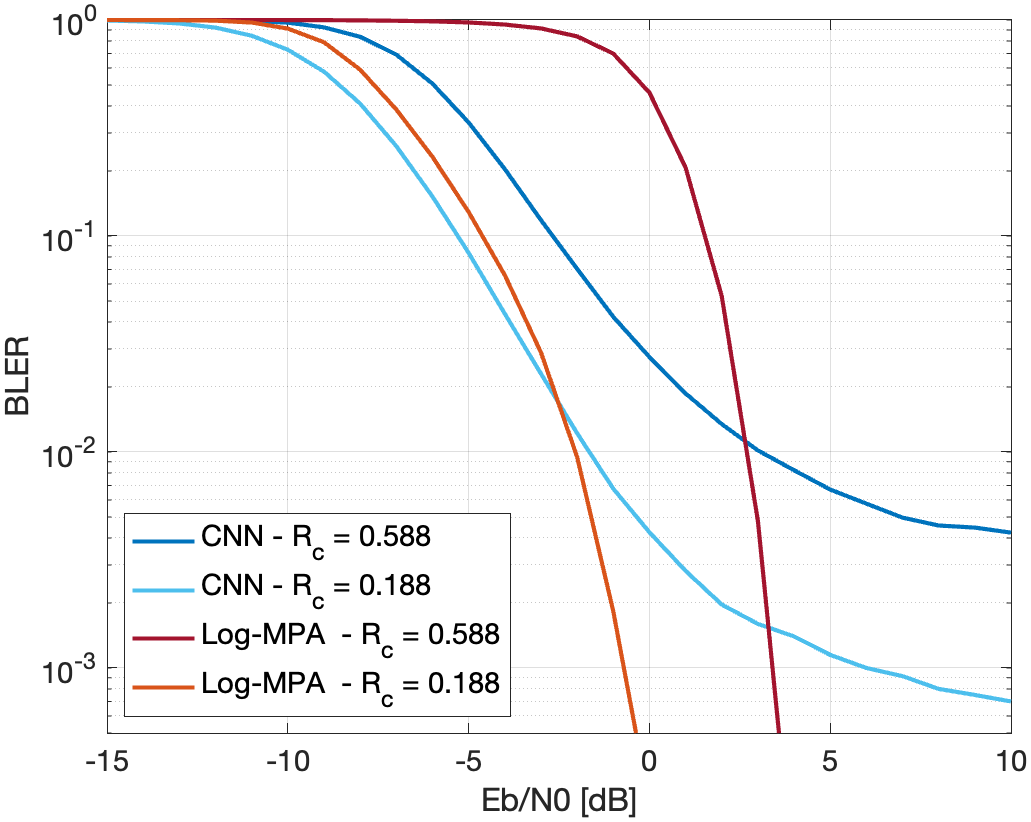}}
    \caption{BLER as a function of the Eb/N0.}
    \label{fig:bler}
\end{figure}
\noindent Figure \ref{fig:bler} reports the BLER achieved by the proposed SCMA receiver as a function of the Eb/N0 computed on the received waveform. Log-MPA with $N_{iter}=10$ iterations is also shown as a benchmark. The plot shows that the AI model provides excellent performance at low Eb/N0 levels. In particular, the CNN achieves a target BLER of 10\% at around -2.75dB of Eb/N0 for $R_c=0.588$ and -5.25dB for $R_c=0.188$. In comparison, Log-MPA requires a 3.25dB and 0.5dB higher Eb/N0 to reach the same BLER target, respectively. The reason for the large gap at high code rate is twofold: firstly, CNNs typically provide good performance in pattern recognition tasks in presence of noise; secondly, Log-MPA introduces approximations to MPA to lower its computational complexity, which in turn leads to suboptimal performance. Due to the training process of the CNN, the layers can approximate the optimal solution with higher accuracy, improving the BLER under low Eb/N0 conditions. Clearly, when robust codes are implemented, more demodulation errors can be corrected, leading to marginal gains with respect to Log-MPA. The BLER performance of the AI-based receiver shows an error floor, resting above $3\cdot10^{-3}$ and $5\cdot10^{-4}$ for high and low code rate, respectively, instead of the classic waterfall trend. As the AI model approximates the optimal solution, residual errors are to be expected even at high Eb/N0 levels, leading to the observed behavior. It must be noted that an increase in the model complexity may result in a lower error floor. Nonetheless, operating at a target 10\% BLER with a relaxed Eb/N0 requirement is a good trade-off for non-critical applications. Indeed, the achieved gain can be exploited by reducing the transmit power of each GD, increasing the devices' expected battery life. While the BLER analysis provides an insight into the error frequency, it fails at highlighting the rate at which information bits are correctly received. From the point of view of the satellite payload, it is of interest to determine the aggregated throughput to be expected at the gNB by the $J$ scheduled GDs. Indeed, apart from the implications from the communications point of view, such quantity also affects computational capabilities requirements for the MEC server. The aggregated theoretical throughput (ATT) can be obtained from the BLER as follows:
\begin{equation}\label{eqn:throughput}
    ATT = \frac{J\cdot N_{TBS}}{T_{TB}}\cdot(1-BLER),
\end{equation}
where $T_{TB}$ represents the duration of the transmission of a TB. Fixing $m=2$ bits per SCMA codeword, and recalling that a codeword spans $K=4$ subcarriers, each GD is able to transmit the entire TB in $N_{sym}=N_{TBS}^{(c)}/(m\cdot K)=12$ symbol times. Thus, considering regular cyclic prefix length and 5G numerology $\mu=0$, each TB is transmitted over one slot, \textit{i.e.}, $T_{TB}=T_{slot}=1ms$. Clearly, the impact of the code rate on Equation \ref{eqn:throughput} is twofold: a lower code rate reduces the BLER, but also decreases the TBS.
\begin{figure}[t]
    \centerline{\includegraphics[width=7.5cm]{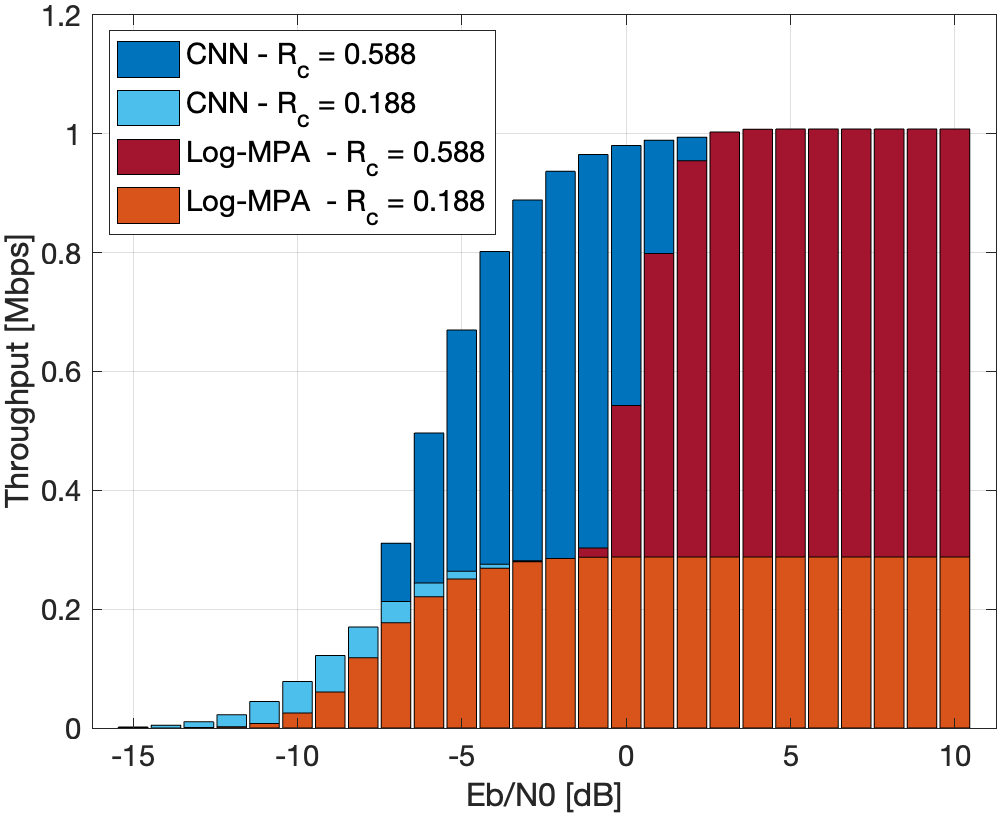}}
    \caption{Aggregated theoretical throughput as a function of the Eb/N0.}
    \label{fig:throughput}
\end{figure}
\noindent The bar plot reported in Figure \ref{fig:throughput} shows the ATT as a function of the Eb/N0 ratio. One can notice that the BLER improvement provided by the AI-based receiver at high code rate result in significant gains in ATT. Indeed, the proposed algorithm provides 800kbps of aggregated throughput at -4dB of Eb/N0, while Log-MPA is not able to operate with such a high noise level, resulting in less than 50kbps of ATT. On the other hand, with $R_c=0.188$ the CNN can provide 170kbps of ATT at -8dB of Eb/N0, a 1dB gain over Log-MPA. Overall, the CNN enables higher code rates to be used at lower Eb/N0 levels, \textit{e.g.}, $R_c=0.588$ at -5dB of Eb/N0 instead of being limited to $R_c=0.188$ with the traditional receiver, resulting in an ATT gain of 420kbps.

\section{Computational complexity}
\noindent Clearly, the CNN-based SCMA receiver can provide considerable communication gains to the MEC NTN system. However, computational complexity is also a key parameter to take into consideration. We compute the number of multiply-and-accumulate (MAC) units as $U_{Conv1D} = L \cdot D \cdot W^{(f)} \cdot N^{(f)}$ for Conv1D layers with $N^{(f)}$ kernels of length $W^{(f)}$ applied to an input matrix of shape $(L, D)$, and $U_{Dense} = N^{(in)} \cdot N^{(out)}$ for Dense layers with $N^{(out)}$ neurons applied to an input vector of length $N^{(in)}$. Considering the model reported in Figure \ref{fig:NNscheme} and the chosen SCMA codebooks, the total amount of MAC operations amounts to 7.9M. On the opposite, recalling the parameters in Table \ref{tab:parameters}, Log-MPA performs 23k multiplications \cite{bib:scmaComplexity}, limiting the number of required MAC units to such value. The increase in MACs by a factor 340 may initially let the reader consider the implementation of the proposed CNN on board of the satellite payload as unfeasible. However, three key considerations should be taken into account. I) Given the vast amount of tunable parameters in AI, it is possible that we did not test specific parameter combinations providing lower computational complexity and comparable, if not higher, link-level performance. II) Complexity reduction and model compression techniques, which are out of the scope of this work, can be applied to greatly reduce the number of MAC units in the proposed model with negligible performance losses or, in some cases, even marginal gains \cite{bib:modelCompression}. III) Specialized hardware accelerators have been successfully implemented on board of power-constrained devices, running inference with large AI models in a timely manner: \textit{e.g.}, in \cite{bib:hardware}, a CNN with 1.1M MACs was implemented on board of a small unmanned aerial vehicle, with the module requiring just 100mW and running at a rate of 139 inferences/s, \textit{i.e.}, 7.19 ms per inference. Hence, the adoption of hardware accelerators, together with model compression techniques and technological advancements, can make the proposed SCMA receiver a feasible solution to enable an efficient massive access in future 6G NTNs.

\section{Conclusions}
\noindent In this work, we applied SCMA to an NTN-based MEC framework to enhance the spectral efficiency of uplink transmissions. To improve the BLER of traditional SCMA receivers at low Eb/N0 levels, where NTN links typically operate, we trained a CNN with MTL to receive SCMA signals. The AI model was evaluated in a link-level simulation, considering both the gNB and the MEC server on board of a LEO satellite. Compared to Log-MPA, the CNN achieves a target 10\% BLER with a gain of up to 3.75dB of Eb/N0. The proposed receiver allows higher code rates to be used at low Eb/N0 levels, resulting in a higher ATT at the MEC server. While the adopted approach is theoretically sound for both NTNs and TNs, its generalization capabilities should be evaluated in future activities. We further assessed the computational complexity of the CNN, concluding that hardware accelerators can enable its on-board implementation. However, SCMA is limited by the channel estimation accuracy over non-orthogonal pilots, an open issue in the SCMA literature. This is exacerbated in LEO-based NTNs due to the large Doppler shift and Doppler rate. Thus, future studies should focus on joint channel estimation and SCMA decoding, possibly considering recurrent NNs. Generative AI approaches could also be investigated to optimize the pilot overhead and the SCMA codebooks in NTNs.

\section{Acknowledgments}\label{Acknowledgment}
\noindent This work was partially supported by the European Union under the Italian National Recovery and Resilience Plan (NRRP) of NextGenerationEU, partnership of "Telecommunications of the Future" (PE00000001 - program "RESTART"), and by the 6G-NTN project, which received funding from the Smart Networks and Services Joint Undertaking (SNS JU) under the European Union’s Horizon Europe research and innovation programme under Grant Agreement No 101096479. The views expressed are those of the authors and do not necessarily represent the project. The Commission is not liable for any use that may be made of any of the information contained therein.

\end{document}